\documentclass[aps,prb,showpacs,twocolumn,superscriptaddress]{revtex4-1}
\usepackage{amsmath}
\usepackage{amsfonts}
\usepackage{braket}
\usepackage{amssymb}
\usepackage{graphicx}
\usepackage{natbib}
\usepackage[colorlinks=true,linkcolor=blue,urlcolor=black,citecolor=blue]{hyperref}
\usepackage{comment}
\usepackage{xcolor}
\definecolor{red}{rgb}{1,0,0}
\definecolor{bostonuniversityred}{rgb}{0.8, 0.0, 0.0}

\usepackage{lipsum}

\allowdisplaybreaks

\begin{document}
\title{{Polaron formation in a spin chain by measurement-induced imaginary Zeeman field}}
\author{P. V. Pyshkin}
\thanks{Corresponding author: pavel.pyshkin@gmail.com}
\affiliation{Department of Physical Chemistry, University of the Basque Country UPV/EHU, 48080 Bilbao, Spain}

\author{E. Ya. Sherman}
\thanks{evgeny.sherman@ehu.eus}
\affiliation{Department of Physical Chemistry, University of the Basque Country UPV/EHU, 48080 Bilbao, Spain}
\affiliation{Ikerbasque, Basque Foundation for Science, 48011 Bilbao, Spain}

\author{Lian-Ao Wu}
\thanks{lianaowu@gmail.com}
\affiliation{Department of Physics, 
University of the Basque Country UPV/EHU, 48080 Bilbao, Spain}
\affiliation{Ikerbasque, Basque Foundation for Science, 48011 Bilbao, Spain}

\date{\today}
\begin{abstract}
{We present a high-rate projective measurement-based approach for controlling non-unitary evolution of a 
quantum chain of interacting spins. In this approach, we demonstrate that local measurement 
of a single external spin coupled to the chain can produce 
a spin polaron, which remains stable after the end of the measurement.} 
This stability results from the fact that the Hilbert space of the chain contains a
subspace of {\em non-decaying} states, stable during the nonunitary evolution. These states
determine the resulting final state of the chain and long-term shape of the polaron.   
In addition to formation of the spin polarons, the presented measurement protocol can be used 
for distillation of non-decaying states from an initial superposition or mixture.  

\end{abstract}
%\pacs{03.65.-w, 42.50.Dv, 37.10.De}
%Quantum mechanics, 03.65.-w
%Quantum state engineering and measurements, 42.50.Dv 
%Cooling of atoms, ions, and molecules, 37.10.De

\maketitle

\section{Introduction}

Control {of evolution} of single-particle- and many-body quantum systems 
is an important branch of modern physics \cite{WisemanMilburn2009,Poggiali2018,Koch2019}
and applied mathematics \cite{Quantum-Control-numerical-book,krotov-book} 
{One of the most interesting realizations of 
such a control is given by the quantum Zeno effect \cite{Itano1990}, where evolution of a quantum 
system undergoing high-rate repeated measurements slows down as a result of the feedback of the 
measurement on the measured system.}  The Zeno effect 
can provide efficient protocols for controlling spin 1/2 in various kinds of measurements and interactions, 
including direct coupling to the environment (e.g., Refs. [\onlinecite{Mundarain2006,Dobrovitski2008,Wolters2013,Luis2013,He2017}]) 
for electrons and spin-orbit coupling \cite{Sherman2014} for cold atoms. 
In addition, the Zeno effect can lead to slow driven spin dynamics 
in quantum dots \cite{Khomitsky2012} and edge magnetization in graphene \cite{Golor2014}. 
Zeno effect plays an important role for electrons coupled to the nuclear 
bath \cite{Klauser2008,Nutz2019,Smirnov2020}. In these systems, 
the finite rate measurements can be used for producing highly polarized states of 
arrays of nuclear spins~\cite{Nucl-Spin-Polar-Measur-Wu2011}. Recently, 
the Zeno effect has been studied in the quantum cavity structures considered as prospective element for 
quantum technologies utilizing light-solid interfaces \cite{Leppenen2021}.

Since interacting quantum spin systems 
demonstrate a rich variety of  phenomena, the understanding of their measurements-based control can provide
protocols useful both for handling quantum information and understanding fundamental aspects of their
physics \cite{Loss1998,Burkard1999,Bose2003,Das2008}. 
Here we study the {physics of repeated projective measurements on a probe
in a system of interacting quantum spins, termed as {\em Zeno-like} effect by 
Refs. [\onlinecite{Hiromichi}] and [\onlinecite{Wu_entanglement_generation}]. 
We consider a single spin coupled to this chain as the probe and show that 
the selective projective measurements on the probe can be used as an instrument 
for producing special, almost stationary, states of the entire quantum system.} These states correspond to spin polarons, the systems
of a broad interest for understanding of the physical properties in various quantum materials 
~\cite{Glazman_Polaron, parton-polaron,polaron-classical-1,polaron-quantum-1,Barabanov-Spin-Polaron}. 
The {effect of these frequent measurements on the probe} amounts to
evolution {of the spin chain} described by a {\em non-Hermitian} Hamiltonian with
a {\em local} imaginary Zeeman field.  The Hilbert space of the system thus consists 
of two subspaces, corresponding to the states, decaying and no-decaying when undergoing the dynamics of the 
{\em non-Hermitian Hamiltonian}. These non-decaying states form a variety of 
spin polarons remaining stable after the termination of the measurement.

This paper is organized as follows. In Sec. II we describe the measured system and the measurement protocol,
derive the corresponding non-Hermitian Hamiltonian, and find the properties and 
dimensionalities of the decaying and non-decaying subspaces. In Sec. III we present numerical results 
of the system evolution during the measurement and demonstrate resulting spin polaron formation.  
%{When the polaron is being formed, the final state of the probed additional spin corresponds to the Zeno effect resulting from the measurement.} 
In addition, in Sec. III 
we demonstrate that the repeated measurement protocol can be used for distillation
of the states in the Hilbert space of interest. Discussions of the results and conclusions are presented in Sec. IV.

\section{Non-unitary evolution of a spin chain}

\subsection{Total Hamiltonian and selective measurements of a probe}

We consider antiferromagnetic chain with $N$ spins $1/2$ described by the Hamiltonian
\begin{equation}\label{H_ch}
	H_{\rm ch} = \sum_{n=1}^{N} \left( X_{n}X_{n+1} + Y_{n}Y_{n+1} \right),
\end{equation}
where the coupling strength is set as as~$1$, and $X_{n}$, $Y_{n}$ are the Pauli matrices for $n$-th spin. 
We use the periodic boundary conditions with $X_{N+n} = X_{n}$ and $Y_{N+n} = Y_{n}.$

\begin{figure}
	\begin{center}
		\includegraphics[width=0.4\textwidth]{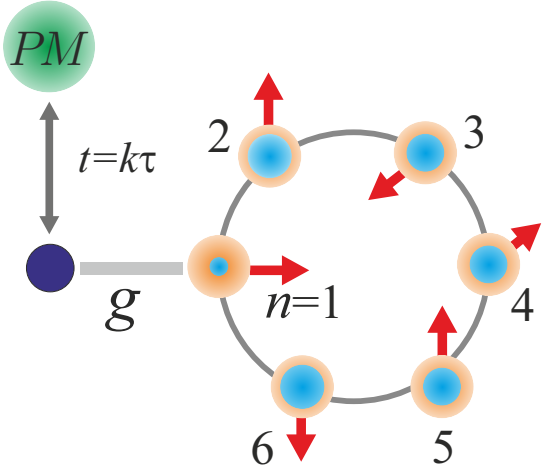}
	\end{center}
	\caption{Schematic picture of a spin chain with locally connected probe (dark circle) being under 
	         frequent selective measurements. The measurements 
	are performed at times $t=k\tau,$ where $k$ is an integer. Red arrows correspond to the in-plane components of the 
	spins and the sizes of the blue circles correspond to their $Z-$components.}
	\label{figchain}
\end{figure}

Now we locally couple an additional probe spin to one of the spins ($n=1$) in the chain, as shown in Fig. \ref{figchain} 
such that the total Hamiltonian of the chain + probe reads
\begin{equation}\label{H_tot}
	H = H_{\rm ch} + g \left(XX_{1} + YY_{1} \right),
\end{equation}
where $X,Y$ without indices refer to the probe, and $g$ is a coupling constant.

We consider the case, where the pure initial state of the total system is $\ket{\Phi_0} = \ket{\uparrow}_{\mathrm{pr}}\otimes\ket{\psi_0}$.
Here $\ket{\uparrow}_{\mathrm{pr}}$ is the state of probe, and $\ket{\psi_0}$ is a state of the chain. 
% Here we assume that  initial state of the chain is pure. 
The total system evolution is governed by Hamiltonian~(\ref{H_tot}). 
After time~$\tau$ we make a projective measurement on the system $\ket{\uparrow}_{\mathrm{pr}}\bra{\uparrow}_{\mathrm{pr}}
\otimes\mathbb{I}_{[N]}$, where $\mathbb{I}_{[N]}$ is the identity operator acting in the Hilbert space of the spin chain. 
If the measurement is successful the state of the system becomes~$\ket{\uparrow}_{\mathrm{pr}} \otimes\ket{\psi_{1}}$, 
where $\ket{\psi_{1}}$ is the evolution-produced state of the chain, and then the next step of evolution and 
measurement can be made. In the opposite case the evolution is discarded. 
After $M$ repetitions the state of the chain becomes

\begin{equation}\label{exact-transform}
	\ket{\psi_0} \rightarrow \ket{\psi_M} = \frac{V^M\ket{\psi_0}}{\sqrt{\braket{\psi_0|V^{\dagger M}V^M|\psi_0 }}},
\end{equation}
where effective non-unitary operator~$V$ acting in the Hilbert space of the spin 
chain defined as $V \equiv \bra{\uparrow}_{\rm pr}\exp(-iH\tau)\ket{\uparrow}_{\rm pr}$ is obtained by taking 
the corresponding matrix element solely in the probe-related subspace. 
The probability of the measurement success at a single step with the number $j>0$ is 
\begin{equation}\label{p_single}
	p_j = \braket{\psi_{j-1}|V^{\dagger}V|\psi_{j-1} }.
\end{equation}
The survival probability of getting~$M$ successful sequential probe measurements is 
\begin{equation}\label{P_M}
	P_M = p_{1}p_2\dots p_M = \braket{\psi_0|V^{\dagger M}V^M|\psi_0 }.
\end{equation}  

We are interested in the {high-rate measurement-produced dynamics}, thus further we assume~$\tau\rightarrow0$. 
The usual feature of such a non-unitary process is a saturation dynamics~\cite{Nucl-Spin-Polar-Measur-Wu2011, Li2011} 
which from the one hand means that $p_k\rightarrow1$ for $k\rightarrow\infty$ (or $P_k\rightarrow{\rm const}>0$), 
and from the other hand the mean value of some operator or a fidelity related to a state of the chain remains a constant 
after a large number of iterations. For example, selective measurements of coupled probe qubit 
can cool a mechanical  oscillator to its ground state ~\cite{Li2011}.

\subsection{Effective Hamiltonian of the spin chain}
In order to qualitatively understand the dynamics of the spin chain and characterize 
the properties of a quantum state after applying infinite number of iterations, we introduce 
the {\em effective} measurement-induced Hamiltonian~$H_{\rm M}$ which acts in the Hilbert space of the spin 
chain such as the non-unitary evolution has the form:   

\begin{equation}\label{Heff-def}
	V = e^{-iH_{\rm M}\tau}.
\end{equation}
Assuming $\tau\rightarrow0$ {(and  $g \tau\rightarrow0$)} we write~$H_{\rm M}$ as a small-$\tau$
expansion with $H_{\rm M}(\tau) = A + \tau B,$ where operators $A$ and $B$ are usually Hermitian and non-Hermitian,
correspondingly. Such a non-Hermitianity has been noticed in earlier publications, for instance, equation 
(21) in Ref. [\onlinecite{Wu_entanglement_generation}]. In order to find $H_{\rm M}$ in (\ref{Heff-def}) 
by matching it with the definition of $V$ following  (\ref{exact-transform}), we write:
\begin{multline}\label{Heff-deriv-1}
	\braket{\uparrow|_{\rm pr}\mathbb{I}_{[N]} - i\tau H - \frac{\tau^2}{2}H^2|\uparrow}_{\rm pr} + \mathcal{O}(\tau^3) = e^{-iA\tau-iB\tau^2}.
\end{multline}
Then, by using explicit Eq. (\ref{H_tot}) we transform (with the accuracy of $\mathcal{O}(\tau^3)$) the LHS of (\ref{Heff-deriv-1}) into
\begin{multline}\label{Heff-deriv-2}
	 \mathbb{I}_{[N]} - i\tau H_{\rm ch} - \frac{\tau^2}{2}H_{\rm ch}^2 - \tau^2 g^2 (\mathbb{I}_{[N]}-Z_{1}\otimes\mathbb{I}_{[N-1]}) = \\
	 e^{-iH_{\rm ch}\tau} - \tau^2 g^2 (\mathbb{I}_{[N]}-Z_{1}\otimes\mathbb{I}_{[N-1]}) =\\
	 e^{-iH_{\rm ch}\tau}e^{-g^2 (\mathbb{I}_{[N]}-Z_{1}\otimes\mathbb{I}_{[N-1]})\tau^2})=\\
	 e^{-iH_{\rm ch}\tau - i (-ig^2 (\mathbb{I}_{[N]}-Z_{1}\otimes\mathbb{I}_{[N-1]}))\tau^2},
\end{multline}
where $Z_{n}$ is the Pauli operator $\sigma_{z}$ acting on the $n$-th spin in the chain and $\mathbb{I}_{[N-1]}$ 
is the identity operator for all the spins except that with $n=1.$
By comparing~(\ref{Heff-deriv-2}) with RHS of~(\ref{Heff-deriv-1}) we 
found $A = H_{\rm ch}$ and $B = -ig^2 (\mathbb{I}_{[N]}-Z_{1}\otimes\mathbb{I}_{[N-1]})$. 
Thus, finally we have
\begin{equation}\label{H_eff_{1}}
	H_{\rm M} = H_{\rm ch} + i g^2 \tau Z_{1}\otimes\mathbb{I}_{[N-1]} - ig^2\tau\mathbb{I}_{[N]}.
\end{equation}  
The form of the second term~(\ref{H_eff_{1}}) shows that our process 
can be mimicked  by introducing an {\em imaginary Zeeman field} \cite{imag-magn-field-1952,imag-magn-field-1967,imag-magn-field-2017}. 
Here we should note: (i) our setting produces a {\em local} imaginary magnetic field instead 
of the uniform one~\cite{non-herm-magnetism-PhysRevX.4.041001} simulated by atomic ensembles with real radiation decay, 
(ii) in the limit~$\tau\rightarrow0$ {(and $g^2\tau\rightarrow0$)} we come 
to the bare spin chain Hamiltonian. The magnitude of this field is proportional to squared coupling 
constant between the probe and the spin from the chain, and linearly proportional 
to the time interval between two measurements. The third term in~(\ref{H_eff_{1}}) ensures 
that the norm of a state vector after acting on the evolution governed by~(\ref{H_eff_{1}}) will not increase.

To clarify the effect of the imaginary magnetic field we rewrite~(\ref{H_eff_{1}}) in the following form:
\begin{equation}\label{H_eff_2}
	H_{\rm M} = H_{\rm ch} - 2i g^2 \tau \Pi_{1}, 
	\quad \Pi_{1} = \ket{\downarrow}_{1}\bra{\downarrow}_{1}\otimes\mathbb{I}_{[N-1]}
\end{equation}
with operator $\Pi_{1}$ acting as a projector $\ket{\downarrow}\bra{\downarrow}$ 
on spin number~$1$ and as unity operator to other spins in chain. 

\subsection{Decaying and non-decaying subspaces}

Dynamical evolution of the initial state~$\ket{\psi_0}$ governed by non-hermitian~$H_{\rm M}$ can 
be described as (see~\cite{non-herm-evol-two-level-quant-sys-PhysRevA.42.1467}):
\begin{equation}\label{non-unitary-evol-1}
\ket{\psi(t)} = \sum_{j=1}^{2^N}\braket{\beta_j|\psi_0}e^{-i\lambda_{j}t}\ket{\alpha_j},	
\end{equation}
where
\begin{equation}\label{eigens-Heff}
 H_{\rm M}\ket{\alpha_j} = \lambda_{j}\ket{\alpha_j}, \; H^{\dagger}_{\rm M}\ket{\beta_j} = \lambda^{*}_{j}\ket{\beta_j}.	
\end{equation}

{The operator $\Pi_1$ in Hamiltonian~(\ref{H_eff_2}) is positive semidefinite, and thus} 
we have~$\mathrm{Im}\lambda_{j}\leq0$, {which leads} to exponential 
suppression of all eigenstates of~$H_{\rm M}$ with~$\mathrm{Im}\lambda_{j}<0$. 
On the contrary, the states with~$\mathrm{Im}\lambda_{j}=0$ will not decay.

The Hilbert space $\mathcal{H}$ of the chain can be divided into two subspaces: 
decaying $\mathcal{H}_{\rm D}$ and non-decaying $\mathcal{H}_{\rm ND}$, 
with $\mathcal{H} = \mathcal{H}_{\rm D}\oplus \mathcal{H}_{\rm ND}$. As a result, any initial state of 
the chain can be decomposed into a superposition of two orthogonal vectors belonging to 
these subspaces: $\ket{\psi_0} = a\ket{\mathrm{D}} + b\ket{\mathrm{ND}}$, 
{where $\ket{\mathrm{D}}$ ($\ket{\mathrm{ND}}$) belongs to the decaying (non-decaying) subspace}. 
The survival probability $P_{\infty}$ of the infinite process is given by: $P_{\infty} = |b|^2$.
The average survival probability for a random pure initial state as well as for thermal mixed 
state with infinite temperature (but not for pure uniform superposition of eigenstates~(\ref{H_ch})) is:
\begin{equation}\label{Prandom}
	\overline{P}_\infty = \frac{\dim\mathcal{H}_{\mathrm{ND}}}{\dim\mathcal{H}}.
\end{equation}
{Here we define the average survival probability as 
$\overline{P}_\infty = 2^{-N}\sum_{m=1}^{2^N}P_\infty(m)$, where $P_\infty(m)$ 
is the survival probability of the infinite process with an initial state 
equal to $m$-th vector of some fixed basis in~$\mathcal{H}$.} 

{Let us define basis~$\{\ket{\phi_j}\}$ with $j=1\dots K$ in~$\mathcal{H}$, where $\ket{\phi_j}$ 
are the eigenstates of~$H_{\rm ch}$ that satisfy
\begin{equation}\label{Hnd_basis_condition}
\Pi_1 \ket{\phi_j} = 0. 
\end{equation}
Note, in the case of degeneracy of~$H_{\rm ch}$ we 
have a freedom to chose~$\{\ket{\phi_j}\}$ to satisfy~(\ref{Hnd_basis_condition}) and thus we can define 
\begin{equation}\label{H_ND}
	\dim\mathcal{H}_{\rm ND} = \max(K).
\end{equation}
Thus from~(\ref{Hnd_basis_condition}) and~(\ref{H_ND}) we conclude 
that $\dim\mathcal{H}_{\mathrm{ND}}$ does not depend on the value of~$g^{2}\tau$ in~(\ref{H_eff_2}).
We find numerically the following expression for non-decaying subspace dimension:
\begin{equation}\label{dimHn}
	\dim\mathcal{H}_{\mathrm{ND}} = \left\{
	\begin{array}{ll}
		2^{(N-1)/{2}}, & \mbox{odd } N\\
		\\
		3\cdot 2^{(N-4)/{2}}, & \mbox{even } N.
	\end{array}
	\right.
\end{equation}
} 

\begin{figure}
	\begin{center}
		\includegraphics{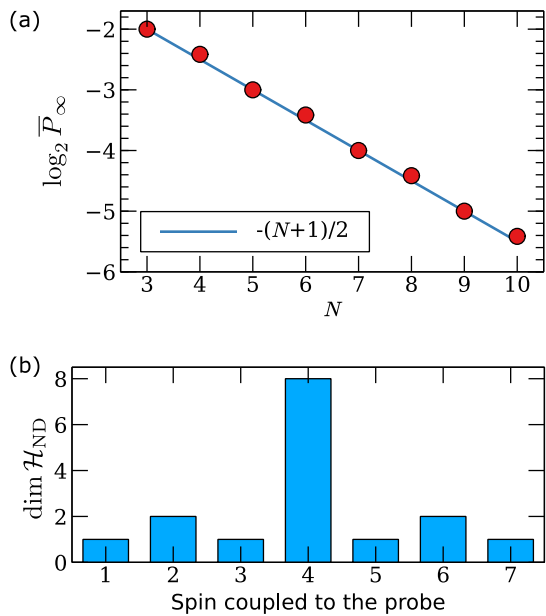}
	\end{center}
	\caption{(a) Numerically estimated survival probability $\overline{P}_\infty$ as a function of spin number~$N$. (b) Dimension of non-decaying subspace versus the number of spin which probe coupled with for open chain with~$N=7$.}
	\label{figdimopen}
\end{figure}

{As it follows from~(\ref{Prandom}) and~(\ref{dimHn}) we have exponential 
decay of~$\overline{P}_\infty$ with increasing of spin chain size~$N$:  
$\overline{P}_\infty \propto 2^{-N/2}$. In Fig. \ref{figdimopen} (a) we show numerical results 
for $\log_2\overline{P}_\infty$ as function of~$N$.} 

{It is interesting to note the following. 
From the one hand~$\dim\mathcal{H}_{\mathrm{ND}}$ is equal to the number of real 
eigenvalues of~(\ref{H_eff_2}), and from the other hand we can make random unitary 
transformations of Hamiltonian~(\ref{H_eff_2}) to obtain a set of random non-Hermitian matrices.
It is known~\cite{real_eigenvalues} that the expected number $R_{L}$ of real eigenvalues 
of an $L\times L$ random non-Hermitian matrix for $L\rightarrow\infty$ is $R_{L} \approx \sqrt{2L/\pi}\propto L^{1/2}$.
In our case the dimension of random matrices is~$L=2^{N}$, thus~$R_{L}\propto2^{N/2}\propto\dim\mathcal{H}_{\mathrm{ND}}$, 
which coincides with~(\ref{dimHn}).}

{For completeness, we mention that for spin chain with open boundary conditions our expression~(\ref{H_eff_2}) 
is the same, except the value of~$\dim\mathcal{H}_{\mathrm{ND}}$ depends on the position of the spin 
connected to the probe. Our numerical estimation shows that~$\dim\mathcal{H}_{\mathrm{ND}}=1$ for any
position of probe for chains with even~$N$, and $\dim\mathcal{H}_{\mathrm{ND}}$ has the maximum value 
for the probe connected to the central spin for chains with odd~$N$ (see Fig.\ref{figdimopen}(b) for example).}

As can be seen from~(\ref{Hnd_basis_condition}) the general form of the state vector which is 
the result of the process after large number of iterations can be written as
\begin{equation}\label{chain-in-infinity}
	\ket{\psi_M} \approx e^{-iH_{\rm M}M\tau} \approx  \ket{\uparrow}_{1}\otimes\ket{\bar{\psi}_M}, \quad M\rightarrow\infty,
\end{equation} 
where $\ket{\bar{\psi}_M}$ is the state of all spins in chain except the first one.
Expression~(\ref{chain-in-infinity}) can be seen as a highly localized 
spin polaron~\cite{Glazman_Polaron, parton-polaron,polaron-classical-1,polaron-quantum-1,Barabanov-Spin-Polaron}.
Note, all states in~$\mathcal{H}_{\rm ND}$ have the form~(\ref{chain-in-infinity}). 
Moreover, the state~$\ket{\uparrow}_{\rm pr}\otimes\ket{{\rm ND}}$ is an eigenstate of 
the total Hamiltonian~(\ref{H_tot}), and this is why it remains stable during the 
process with survival probability of each step equal to~$1$, %{corresponding to the conventional Zeno effect on the measured system.}
In other words, the 
probe remains disentangled from the chain being in~$\ket{{\rm ND}}$ state under 
the evolution governed by~(\ref{H_tot}). This situation may be comparable with the 
chain state being in the decoherence free subspace~\cite{Wu2005}.

\section{Numerical results for spin polaron formation}
\begin{figure}
	\begin{center}
		\includegraphics{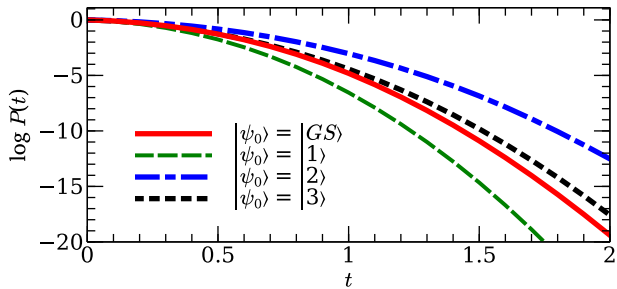}
	\end{center}
	\caption{Survival probability $P$ as a function of time for the first four 
	eigenstates as initial states of the spin chain with~$N=6$, $\tau=0.03$ and $g=4$.}
	\label{figPtot1}
\end{figure}

To numerically illustrate the above statements on the decaying and non-decaying subspaces
and demonstrate the evolution 
of the measured system, we use an $N=6$ spin chain, with coupling to 
probe with strength~$g=4$, and intervals between measurements are~$\tau=0.03$.
We numerically found that for chosen spin chain first eigenstates are 
lying in~the $\mathcal{H}_{\rm D}$ subspace. 
In Fig.\ref{figPtot1} we show how survival probability decreases with time for initial states 
from the ~$\mathcal{H}_{\rm D}$ subspace. 

\begin{figure}
	\begin{center}
		\includegraphics{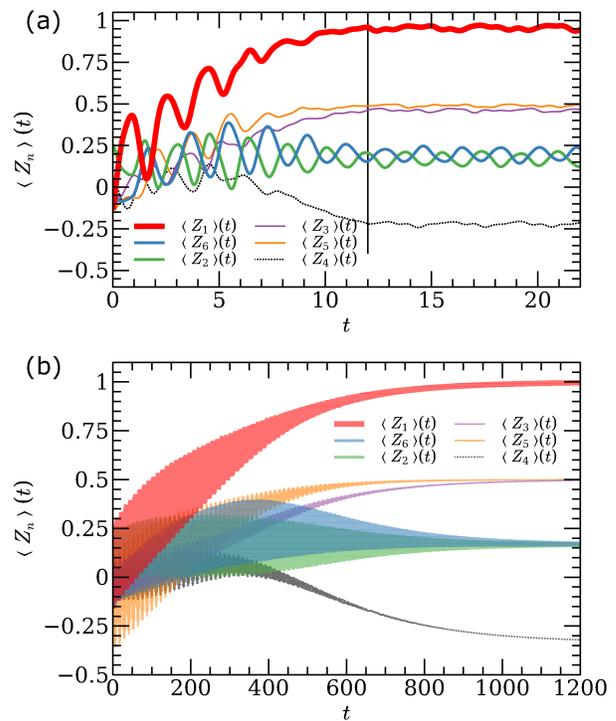}
	\end{center}
	\caption{Example of magnetization dynamics for the spin chain with~$N=6$, $\tau=0.03$. 
	Panel (a):  $g=4$, the probe spin disconnected at $t=12$ (vertical line). 
	Panel (b): $g=0.5$, the probe is not disconnected. Numeration of spins corresponds to Fig. \ref{figchain}.}
	\label{figsingleexample}
\end{figure}

As an example of dynamical spin polaron 
formation we use initial state~$\ket{\psi_0}=\sum_{l=0}^{16}\ket{l}/\sqrt{17}$ 
which is the one of possible uniform superposition of eigenstates of 
the first $4$ energy levels of the chain with~$N=6$, $g=4$ and $\tau=0.03$. 
In this setting the groundstate is non-degenerate, the first excited level 
has degeneracy~$2$, the second level has degeneracy $6$, and the third one has the degeneracy $8$. In Fig.~\ref{figsingleexample}(a) 
we show magnetization~$\braket{\psi_k|Z_{n}|\psi_k}$ of each spin in chain, where $k$ is 
a number of step. For convenience we use time $t=k\tau$ instead of number $k$ in the plot. 
Interaction and measurements with the probe are switched off at~$t=12$ (corresponding to $M=400$ steps), 
and after this spin chain evolution is governed by the Hamiltonian~(\ref{H_ch}). 
Survival probability of such a process (for our choosing of 
eigenstates in degenerate subspaces) is~$P_{400}\approx 0.02$. 
{For comparison, in Fig.~\ref{figsingleexample}(b) we let~$g=0.5$ and the non-unitary process is not interrupted. 
The survival probability of this process~$P_\infty\approx0.017$, which coincides with squared amplitude 
of being initial state in the~$\mathcal{H}_{\rm ND}$. Note that, as expected, $\mathcal{H}_{\rm ND}$ 
is the same for both $g=4$ and $g=0.5$.}
Figure ~\ref{figsingleexample} shows that the process ends up with 
the formation of a localized spin polaron at spin~$1$ of the chain. As can be seen, 
the total magnetization of the chain increases during the non-unitary process, 
and after switching off the probe remains constant due to the fact $[\sum_{n}Z_{n}, H_{\rm ch}]=0$.

\begin{figure}
	\begin{center}
		\includegraphics{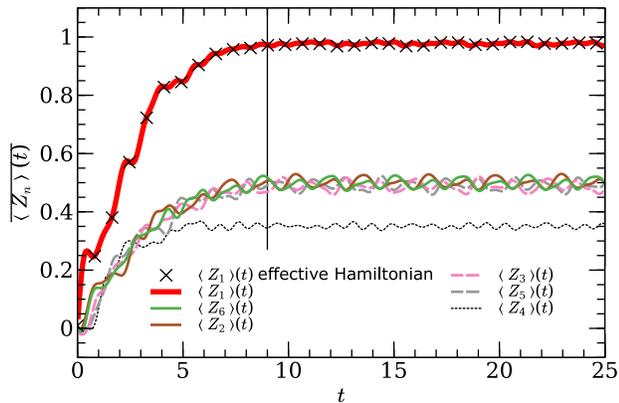}
	\end{center}
	\caption{Averaged local magnetization from the ensemble of $100$ random initial states. 
	The probe spin disconnected at $t=9$ (vertical line) and after this time spin chain evolution is governed by Hamiltonian~(\ref{H_ch}).}
	\label{figaveraged100}
\end{figure}

To study averaged dynamics we use $100$ randomly chosen initial states~$\ket{\psi_0}$ of the 
chain and plot average magnetization for this ensemble in Fig.~\ref{figaveraged100}.
In this simulation we apply $M=300$ evolution-measurement cycles, i.e. after time $T = 300\tau = 9$ 
(vertical line in Fig.~\ref{figaveraged100}) the evolution of the spin chain is governed by 
the Hamiltonian~(\ref{H_ch}) and probe is disconnected from the chain. 
As can be seen in Fig. \ref{figaveraged100}, quantum state of each chain in the 
ensemble approximately satisfies~(\ref{chain-in-infinity}) as a result of the non-unitary process. 
The relation (\ref{chain-in-infinity}) also holds after switching off the probe. 
This is due to our special choice of couplings inside the chain, and between the chain and probe: 
the first term in RHS of~(\ref{H_eff_2}) coincides with the bare spin chain Hamiltonian~(\ref{H_ch}). 
Also we can see that for our choice of $\tau$ and $g,$ the result from a direct simulation of our discrete 
algorithm coincides with the result obtained from continuous evolution governed by non-Hermitian 
Hamiltonian~(\ref{H_eff_2}). The average survival probability of the whole process is~$P_{300}\approx 0.097$ 
which is little higher than~$6/64\approx0.94$ predicted by expression~(\ref{Prandom}) with $\dim\mathcal{H}_{\rm ND}=6$ 
and $\dim\mathcal{H}=64$ due to $P_\infty<P_{300}$.

\begin{figure}
	\begin{center}
		\includegraphics{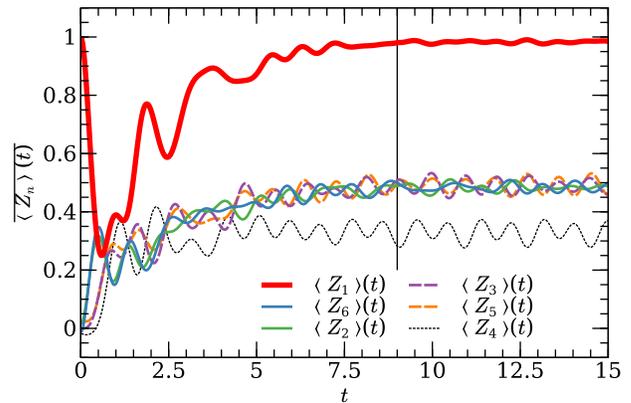}
	\end{center}
	\caption{Averaged local magnetization from the ensemble of 100 random initial states in 
	the form $\ket{\psi_0} = \ket{\uparrow}_{1}\otimes\ket{\mathrm{Random\, state}}$.}
	\label{figaveragedpolyaron100}
\end{figure}

The relation (\ref{chain-in-infinity}) is necessary but not sufficient for 
a state~$\ket{\psi}$ to belong to $\mathcal{H}_{\rm ND}$. 
To illustrate this circumstance, we show in Fig.~\ref{figaveragedpolyaron100} the averaged result of $100$ processes with initial state 
\begin{equation}\label{init_polaron}
\ket{\psi_0} = \ket{\uparrow}_{1}\otimes\ket{\mathrm{Random\, state}}.	
\end{equation}
As can be seen in Fig. \ref{figaveragedpolyaron100}, such a choice of initial state can be treated as an 
unstable polaron which, in turn, can be transformed into a stable one via a non-unitary process.
The average survival probability for such a special initial state in our simulation is~$P_{300}\approx0.19$ which 
is approximately in a factor of two larger than that for totally random initial states~(\ref{Prandom}) in the previous example. 
This is because when we choose the initial state in a form~(\ref{init_polaron}) we effectively 
truncate~$\dim\mathcal{H}$ in expression~(\ref{Prandom}) twice and do not change~$\dim\mathcal{H}_{\rm ND}$.

Exponential sensitivity to initial states is a characteristic feature~\cite{exponential-sens,Measur_indused_orthogonalization} of  
non-unitary processes in quantum mechanics. Described here algorithm can be used to extract states
being in the~$\mathcal{H}_{\rm ND}$ from the initial superposition even for the cases of a small amplitude of such states.
The ``price'' for this possibility is proportionally small survival probability~$P_\infty$. 
As can be easily found the ground state~$\ket{\mathrm{GS}}$ of a spin chain~(\ref{H_ch}) with~$N=6$ 
is a decaying state, while the state~$\ket{\Uparrow}=\ket{\uparrow\uparrow\dots\uparrow}$ with all spins 
being up is a non-decaying one.

\begin{figure}
	\begin{center}
		\includegraphics{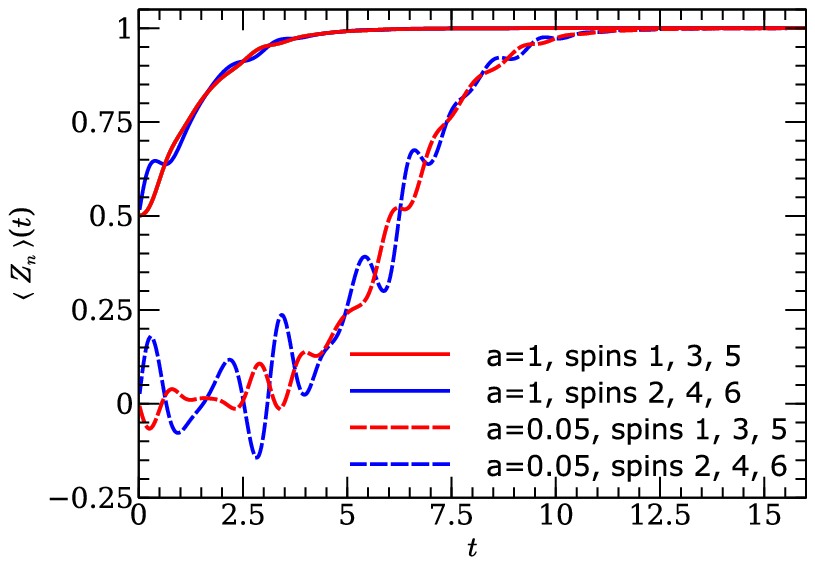}
	\end{center}
	\caption{Distillation of an non-decaying state $\ket{\Uparrow}$ 
	from initial superposition $\ket{\psi_0} = (\ket{\mathrm{GS}} + a\ket{\Uparrow})(1+|a|^2)^{-1/2}$  }
	\label{figdistill}
\end{figure}

Let us take initial state in the form~$\ket{\psi_0} = (\ket{\mathrm{GS}} + a\ket{\Uparrow})(1+|a|^2)^{-1/2}$, 
and simulate the process with~$g=4$ and $\tau=0.03$ for $a=1$ and $a=0.05$ (see Fig.\ref{figdistill}). 
As expected, the survival probability is $P_\infty = |a|^2 / (1 + |a|^2)$ and as a result we 
have a chain in the state~$\ket{\Uparrow}$. Note, in this example we achieve not a polaron but a 
uniform magnetization of a chain by using local control. Exponential amplification of such  
small perturbations in the initial states also can be seen as a ``butterfly effect''.

Note that if the initial state of the spin chain is a mixture of the form
\begin{equation}
\rho_0 = p\ket{\mathrm{ND}}\bra{\mathrm{ND}} + (1-p)\ket{\mathrm{D}}\bra{\mathrm{D}},	
\end{equation}
our algorithm transforms it to a pure state~$\ket{\mathrm{ND}}$ with survival probability $P_\infty=p$. 
Thus our proposal can distill one many-body state from another via 
only local~\cite{Pyshkin_Spin_cutting_NJP, Spin_Chain_local_cut} control.

{As mentioned in the above, the proposed process can also be used for 
quantum state discrimination~\cite{discrimination_Bae_2015}. 
For instance, if we have an {\em a priori} known set of 
states $\{ \ket{\psi_1}=\ket{\mathrm{ND}} ,\; \ket{\psi_2} = a \ket{\mathrm{ND}} + b \ket{\mathrm{D}} \}$, 
and initial state of the chain is one of them, we can distinguish these two possibilities 
via running our non-unitary process and check if outcome~$\ket{\downarrow}_{\rm pr}$ happens. 
In general case $|\braket{\psi_1|\psi_2}|\geq 0$, and we need only one copy of a quantum state.}

\section{Discussion and conclusion}   

We have shown that the {effect of high-rate projective measurements}
of a single spin coupled to an antiferromagnetic 
quantum spin $1/2$ chain can be mimicked by the evolution {of the chain} caused by a non-Hermitian Hamiltonian.
This Hamiltonian includes an imaginary Zeeman field determined by a short time 
interval between the measurements and coupling between the measured spin and the chain. 
The physical origin of this non-Hermitianity, not based on the decay 
of the states involved in the spin dynamics and measurement, is, therefore,
qualitatively different from that presented in Ref. [\onlinecite{non-herm-magnetism-PhysRevX.4.041001}]. 
We obtained this non-Hermitian operator and 
identified the sets of states and corresponding Hilbert subspaces,
decaying and surviving under the action of this measurement-produced Hamiltonian. 
Thus, the measurement process results in the distillation of the corresponding states. The surviving states 
produce a stable ferromagnetic (with nonzero total spin) polaron with the expectation values of 
the spins oscillating around some stationary values after the end of the measurement. 
Since the produced polarons depend on the initial state of the quantum chain and the measurement protocol, 
these results can be applied to formation of quantum spin states on demand as well.

\section*{Acknowledgments}
We acknowledge support of the Spanish Ministry of Science and the European Regional Development
Fund through PGC2018-101355-B-I00 (MCIU/AEI/FEDER, UE), and the Basque Country Government through
Grant No. IT986-16.

\bibliographystyle{apsrev4-1}
%\bibliography{biblioteka}

\begin{thebibliography}{43}%
	\makeatletter
	\providecommand \@ifxundefined [1]{%
		\@ifx{#1\undefined}
	}%
	\providecommand \@ifnum [1]{%
		\ifnum #1\expandafter \@firstoftwo
		\else \expandafter \@secondoftwo
		\fi
	}%
	\providecommand \@ifx [1]{%
		\ifx #1\expandafter \@firstoftwo
		\else \expandafter \@secondoftwo
		\fi
	}%
	\providecommand \natexlab [1]{#1}%
	\providecommand \enquote  [1]{``#1''}%
	\providecommand \bibnamefont  [1]{#1}%
	\providecommand \bibfnamefont [1]{#1}%
	\providecommand \citenamefont [1]{#1}%
	\providecommand \href@noop [0]{\@secondoftwo}%
	\providecommand \href [0]{\begingroup \@sanitize@url \@href}%
	\providecommand \@href[1]{\@@startlink{#1}\@@href}%
	\providecommand \@@href[1]{\endgroup#1\@@endlink}%
	\providecommand \@sanitize@url [0]{\catcode `\\12\catcode `\$12\catcode
		`\&12\catcode `\#12\catcode `\^12\catcode `\_12\catcode `\%12\relax}%
	\providecommand \@@startlink[1]{}%
	\providecommand \@@endlink[0]{}%
	\providecommand \url  [0]{\begingroup\@sanitize@url \@url }%
	\providecommand \@url [1]{\endgroup\@href {#1}{\urlprefix }}%
	\providecommand \urlprefix  [0]{URL }%
	\providecommand \Eprint [0]{\href }%
	\providecommand \doibase [0]{http://dx.doi.org/}%
	\providecommand \selectlanguage [0]{\@gobble}%
	\providecommand \bibinfo  [0]{\@secondoftwo}%
	\providecommand \bibfield  [0]{\@secondoftwo}%
	\providecommand \translation [1]{[#1]}%
	\providecommand \BibitemOpen [0]{}%
	\providecommand \bibitemStop [0]{}%
	\providecommand \bibitemNoStop [0]{.\EOS\space}%
	\providecommand \EOS [0]{\spacefactor3000\relax}%
	\providecommand \BibitemShut  [1]{\csname bibitem#1\endcsname}%
	\let\auto@bib@innerbib\@empty
	%</preamble>
	\bibitem [{\citenamefont {Wiseman}\ and\ \citenamefont
		{Milburn}(2009)}]{WisemanMilburn2009}%
	\BibitemOpen
	\bibfield  {author} {\bibinfo {author} {\bibfnamefont {H.~M.}\ \bibnamefont
			{Wiseman}}\ and\ \bibinfo {author} {\bibfnamefont {G.~J.}\ \bibnamefont
			{Milburn}},\ }\href@noop {} {\emph {\bibinfo {title} {Quantum Measurement and
				Control}}}\ (\bibinfo  {publisher} {Cambridge University Press},\ \bibinfo
	{year} {2009})\BibitemShut {NoStop}%
	\bibitem [{\citenamefont {Poggiali}\ \emph {et~al.}(2018)\citenamefont
		{Poggiali}, \citenamefont {Cappellaro},\ and\ \citenamefont
		{Fabbri}}]{Poggiali2018}%
	\BibitemOpen
	\bibfield  {author} {\bibinfo {author} {\bibfnamefont {F.}~\bibnamefont
			{Poggiali}}, \bibinfo {author} {\bibfnamefont {P.}~\bibnamefont
			{Cappellaro}}, \ and\ \bibinfo {author} {\bibfnamefont {N.}~\bibnamefont
			{Fabbri}},\ }\href@noop {} {\bibfield  {journal} {\bibinfo  {journal}
			{Physical Review X}\ }\textbf {\bibinfo {volume} {8}},\ \bibinfo {pages}
		{021059} (\bibinfo {year} {2018})}\BibitemShut {NoStop}%
	\bibitem [{\citenamefont {Koch}\ \emph {et~al.}(2019)\citenamefont {Koch},
		\citenamefont {Lemeshko},\ and\ \citenamefont {Sugny}}]{Koch2019}%
	\BibitemOpen
	\bibfield  {author} {\bibinfo {author} {\bibfnamefont {C.~P.}\ \bibnamefont
			{Koch}}, \bibinfo {author} {\bibfnamefont {M.}~\bibnamefont {Lemeshko}}, \
		and\ \bibinfo {author} {\bibfnamefont {D.}~\bibnamefont {Sugny}},\
	}\href@noop {} {\bibfield  {journal} {\bibinfo  {journal} {Reviews of Modern
				Physics}\ }\textbf {\bibinfo {volume} {91}},\ \bibinfo {pages} {035005}
		(\bibinfo {year} {2019})}\BibitemShut {NoStop}%
	\bibitem [{\citenamefont {Borzi}\ \emph {et~al.}(2017)\citenamefont {Borzi},
		\citenamefont {Ciaramella},\ and\ \citenamefont
		{Sprengel}}]{Quantum-Control-numerical-book}%
	\BibitemOpen
	\bibfield  {author} {\bibinfo {author} {\bibfnamefont {A.}~\bibnamefont
			{Borzi}}, \bibinfo {author} {\bibfnamefont {G.}~\bibnamefont {Ciaramella}}, \
		and\ \bibinfo {author} {\bibfnamefont {M.}~\bibnamefont {Sprengel}},\
	}\href@noop {} {\emph {\bibinfo {title} {Formulation and Numerical Solution
				of Quantum Control Problems (Computational Science \& Engineering)}}}\
	(\bibinfo  {publisher} {SIAM-Society for Industrial \& Applied Mathematics},\
	\bibinfo {year} {2017})\BibitemShut {NoStop}%
	\bibitem [{\citenamefont {Krotov}(1995)}]{krotov-book}%
	\BibitemOpen
	\bibfield  {author} {\bibinfo {author} {\bibfnamefont {V.}~\bibnamefont
			{Krotov}},\ }\href@noop {} {\emph {\bibinfo {title} {Global Methods in
				Optimal Control Theory (Chapman \& Hall/CRC Pure and Applied Mathematics)}}}\
	(\bibinfo  {publisher} {CRC Press},\ \bibinfo {year} {1995})\BibitemShut
	{NoStop}%
	\bibitem [{\citenamefont {Itano}\ \emph {et~al.}(1990)\citenamefont {Itano},
		\citenamefont {Heinzen}, \citenamefont {Bollinger},\ and\ \citenamefont
		{Wineland}}]{Itano1990}%
	\BibitemOpen
	\bibfield  {author} {\bibinfo {author} {\bibfnamefont {W.~M.}\ \bibnamefont
			{Itano}}, \bibinfo {author} {\bibfnamefont {D.~J.}\ \bibnamefont {Heinzen}},
		\bibinfo {author} {\bibfnamefont {J.~J.}\ \bibnamefont {Bollinger}}, \ and\
		\bibinfo {author} {\bibfnamefont {D.~J.}\ \bibnamefont {Wineland}},\ }\href
	{\doibase 10.1103/PhysRevA.41.2295} {\bibfield  {journal} {\bibinfo
			{journal} {Phys. Rev. A}\ }\textbf {\bibinfo {volume} {41}},\ \bibinfo
		{pages} {2295} (\bibinfo {year} {1990})}\BibitemShut {NoStop}%
	\bibitem [{\citenamefont {Mundarain}\ and\ \citenamefont
		{Stephany}(2006)}]{Mundarain2006}%
	\BibitemOpen
	\bibfield  {author} {\bibinfo {author} {\bibfnamefont {D.~F.}\ \bibnamefont
			{Mundarain}}\ and\ \bibinfo {author} {\bibfnamefont {J.}~\bibnamefont
			{Stephany}},\ }\href {\doibase 10.1103/PhysRevA.73.042113} {\bibfield
		{journal} {\bibinfo  {journal} {Phys. Rev. A}\ }\textbf {\bibinfo {volume}
			{73}},\ \bibinfo {pages} {042113} (\bibinfo {year} {2006})}\BibitemShut
	{NoStop}%
	\bibitem [{\citenamefont {Dobrovitski}\ \emph {et~al.}(2008)\citenamefont
		{Dobrovitski}, \citenamefont {Feiguin}, \citenamefont {Awschalom},\ and\
		\citenamefont {Hanson}}]{Dobrovitski2008}%
	\BibitemOpen
	\bibfield  {author} {\bibinfo {author} {\bibfnamefont {V.~V.}\ \bibnamefont
			{Dobrovitski}}, \bibinfo {author} {\bibfnamefont {A.~E.}\ \bibnamefont
			{Feiguin}}, \bibinfo {author} {\bibfnamefont {D.~D.}\ \bibnamefont
			{Awschalom}}, \ and\ \bibinfo {author} {\bibfnamefont {R.}~\bibnamefont
			{Hanson}},\ }\href {\doibase 10.1103/PhysRevB.77.245212} {\bibfield
		{journal} {\bibinfo  {journal} {Phys. Rev. B}\ }\textbf {\bibinfo {volume}
			{77}},\ \bibinfo {pages} {245212} (\bibinfo {year} {2008})}\BibitemShut
	{NoStop}%
	\bibitem [{\citenamefont {Wolters}\ \emph {et~al.}(2013)\citenamefont
		{Wolters}, \citenamefont {Strau\ss{}}, \citenamefont {Schoenfeld},\ and\
		\citenamefont {Benson}}]{Wolters2013}%
	\BibitemOpen
	\bibfield  {author} {\bibinfo {author} {\bibfnamefont {J.}~\bibnamefont
			{Wolters}}, \bibinfo {author} {\bibfnamefont {M.}~\bibnamefont {Strau\ss{}}},
		\bibinfo {author} {\bibfnamefont {R.~S.}\ \bibnamefont {Schoenfeld}}, \ and\
		\bibinfo {author} {\bibfnamefont {O.}~\bibnamefont {Benson}},\ }\href
	{\doibase 10.1103/PhysRevA.88.020101} {\bibfield  {journal} {\bibinfo
			{journal} {Phys. Rev. A}\ }\textbf {\bibinfo {volume} {88}},\ \bibinfo
		{pages} {020101(R)} (\bibinfo {year} {2013})}\BibitemShut {NoStop}%
	\bibitem [{\citenamefont {Luis}\ \emph {et~al.}(2013)\citenamefont {Luis},
		\citenamefont {Gonzalo},\ and\ \citenamefont {Porras}}]{Luis2013}%
	\BibitemOpen
	\bibfield  {author} {\bibinfo {author} {\bibfnamefont {A.}~\bibnamefont
			{Luis}}, \bibinfo {author} {\bibfnamefont {I.}~\bibnamefont {Gonzalo}}, \
		and\ \bibinfo {author} {\bibfnamefont {M.~A.}\ \bibnamefont {Porras}},\
	}\href {\doibase 10.1103/PhysRevA.87.064102} {\bibfield  {journal} {\bibinfo
			{journal} {Phys. Rev. A}\ }\textbf {\bibinfo {volume} {87}},\ \bibinfo
		{pages} {064102} (\bibinfo {year} {2013})}\BibitemShut {NoStop}%
	\bibitem [{\citenamefont {He}\ \emph {et~al.}(2017)\citenamefont {He},
		\citenamefont {Chen},\ and\ \citenamefont {Zheng}}]{He2017}%
	\BibitemOpen
	\bibfield  {author} {\bibinfo {author} {\bibfnamefont {S.}~\bibnamefont
			{He}}, \bibinfo {author} {\bibfnamefont {Q.-H.}\ \bibnamefont {Chen}}, \ and\
		\bibinfo {author} {\bibfnamefont {H.}~\bibnamefont {Zheng}},\ }\href
	{\doibase 10.1103/PhysRevA.95.062109} {\bibfield  {journal} {\bibinfo
			{journal} {Phys. Rev. A}\ }\textbf {\bibinfo {volume} {95}},\ \bibinfo
		{pages} {062109} (\bibinfo {year} {2017})}\BibitemShut {NoStop}%
	\bibitem [{\citenamefont {Sherman}\ and\ \citenamefont
		{Sokolovski}(2014)}]{Sherman2014}%
	\BibitemOpen
	\bibfield  {author} {\bibinfo {author} {\bibfnamefont {E.~Y.}\ \bibnamefont
			{Sherman}}\ and\ \bibinfo {author} {\bibfnamefont {D.}~\bibnamefont
			{Sokolovski}},\ }\href {\doibase 10.1088/1367-2630/16/1/015013} {\bibfield
		{journal} {\bibinfo  {journal} {New Journal of Physics}\ }\textbf {\bibinfo
			{volume} {16}},\ \bibinfo {pages} {015013} (\bibinfo {year}
		{2014})}\BibitemShut {NoStop}%
	\bibitem [{\citenamefont {Khomitsky}\ \emph {et~al.}(2012)\citenamefont
		{Khomitsky}, \citenamefont {Gulyaev},\ and\ \citenamefont
		{Sherman}}]{Khomitsky2012}%
	\BibitemOpen
	\bibfield  {author} {\bibinfo {author} {\bibfnamefont {D.~V.}\ \bibnamefont
			{Khomitsky}}, \bibinfo {author} {\bibfnamefont {L.~V.}\ \bibnamefont
			{Gulyaev}}, \ and\ \bibinfo {author} {\bibfnamefont {E.~Y.}\ \bibnamefont
			{Sherman}},\ }\href {\doibase 10.1103/PhysRevB.85.125312} {\bibfield
		{journal} {\bibinfo  {journal} {Phys. Rev. B}\ }\textbf {\bibinfo {volume}
			{85}},\ \bibinfo {pages} {125312} (\bibinfo {year} {2012})}\BibitemShut
	{NoStop}%
	\bibitem [{\citenamefont {Golor}\ \emph {et~al.}(2014)\citenamefont {Golor},
		\citenamefont {Wessel},\ and\ \citenamefont {Schmidt}}]{Golor2014}%
	\BibitemOpen
	\bibfield  {author} {\bibinfo {author} {\bibfnamefont {M.}~\bibnamefont
			{Golor}}, \bibinfo {author} {\bibfnamefont {S.}~\bibnamefont {Wessel}}, \
		and\ \bibinfo {author} {\bibfnamefont {M.~J.}\ \bibnamefont {Schmidt}},\
	}\href {\doibase 10.1103/PhysRevLett.112.046601} {\bibfield  {journal}
		{\bibinfo  {journal} {Phys. Rev. Lett.}\ }\textbf {\bibinfo {volume} {112}},\
		\bibinfo {pages} {046601} (\bibinfo {year} {2014})}\BibitemShut {NoStop}%
	\bibitem [{\citenamefont {Klauser}\ \emph {et~al.}(2008)\citenamefont
		{Klauser}, \citenamefont {Coish},\ and\ \citenamefont {Loss}}]{Klauser2008}%
	\BibitemOpen
	\bibfield  {author} {\bibinfo {author} {\bibfnamefont {D.}~\bibnamefont
			{Klauser}}, \bibinfo {author} {\bibfnamefont {W.~A.}\ \bibnamefont {Coish}},
		\ and\ \bibinfo {author} {\bibfnamefont {D.}~\bibnamefont {Loss}},\ }\href
	{\doibase 10.1103/PhysRevB.78.205301} {\bibfield  {journal} {\bibinfo
			{journal} {Phys. Rev. B}\ }\textbf {\bibinfo {volume} {78}},\ \bibinfo
		{pages} {205301} (\bibinfo {year} {2008})}\BibitemShut {NoStop}%
	\bibitem [{\citenamefont {Nutz}\ \emph {et~al.}(2019)\citenamefont {Nutz},
		\citenamefont {Androvitsaneas}, \citenamefont {Young}, \citenamefont
		{Oulton},\ and\ \citenamefont {McCutcheon}}]{Nutz2019}%
	\BibitemOpen
	\bibfield  {author} {\bibinfo {author} {\bibfnamefont {T.}~\bibnamefont
			{Nutz}}, \bibinfo {author} {\bibfnamefont {P.}~\bibnamefont
			{Androvitsaneas}}, \bibinfo {author} {\bibfnamefont {A.}~\bibnamefont
			{Young}}, \bibinfo {author} {\bibfnamefont {R.}~\bibnamefont {Oulton}}, \
		and\ \bibinfo {author} {\bibfnamefont {D.~P.~S.}\ \bibnamefont
			{McCutcheon}},\ }\href {\doibase 10.1103/PhysRevA.99.053853} {\bibfield
		{journal} {\bibinfo  {journal} {Phys. Rev. A}\ }\textbf {\bibinfo {volume}
			{99}},\ \bibinfo {pages} {053853} (\bibinfo {year} {2019})}\BibitemShut
	{NoStop}%
	\bibitem [{\citenamefont {Smirnov}\ \emph {et~al.}(2020)\citenamefont
		{Smirnov}, \citenamefont {Zhukov}, \citenamefont {Yakovlev}, \citenamefont
		{Kirstein}, \citenamefont {Bayer},\ and\ \citenamefont
		{Greilich}}]{Smirnov2020}%
	\BibitemOpen
	\bibfield  {author} {\bibinfo {author} {\bibfnamefont {D.~S.}\ \bibnamefont
			{Smirnov}}, \bibinfo {author} {\bibfnamefont {E.~A.}\ \bibnamefont {Zhukov}},
		\bibinfo {author} {\bibfnamefont {D.~R.}\ \bibnamefont {Yakovlev}}, \bibinfo
		{author} {\bibfnamefont {E.}~\bibnamefont {Kirstein}}, \bibinfo {author}
		{\bibfnamefont {M.}~\bibnamefont {Bayer}}, \ and\ \bibinfo {author}
		{\bibfnamefont {A.}~\bibnamefont {Greilich}},\ }\href {\doibase
		10.1103/PhysRevB.102.235413} {\bibfield  {journal} {\bibinfo  {journal}
			{Phys. Rev. B}\ }\textbf {\bibinfo {volume} {102}},\ \bibinfo {pages}
		{235413} (\bibinfo {year} {2020})}\BibitemShut {NoStop}%
	\bibitem [{\citenamefont {Wu}(2011)}]{Nucl-Spin-Polar-Measur-Wu2011}%
	\BibitemOpen
	\bibfield  {author} {\bibinfo {author} {\bibfnamefont {L.-A.}\ \bibnamefont
			{Wu}},\ }\href@noop {} {\bibfield  {journal} {\bibinfo  {journal} {Journal of
				Physics A: Mathematical and Theoretical}\ }\textbf {\bibinfo {volume} {44}},\
		\bibinfo {pages} {325302} (\bibinfo {year} {2011})}\BibitemShut {NoStop}%
	\bibitem [{\citenamefont {Leppenen}\ \emph {et~al.}(2021)\citenamefont
		{Leppenen}, \citenamefont {Lanco},\ and\ \citenamefont
		{Smirnov}}]{Leppenen2021}%
	\BibitemOpen
	\bibfield  {author} {\bibinfo {author} {\bibfnamefont {N.~V.}\ \bibnamefont
			{Leppenen}}, \bibinfo {author} {\bibfnamefont {L.}~\bibnamefont {Lanco}}, \
		and\ \bibinfo {author} {\bibfnamefont {D.~S.}\ \bibnamefont {Smirnov}},\
	}\href {\doibase 10.1103/physrevb.103.045413} {\bibfield  {journal} {\bibinfo
			{journal} {Physical Review B}\ }\textbf {\bibinfo {volume} {103}},\ \bibinfo
		{pages} {045413} (\bibinfo {year} {2021})}\BibitemShut {NoStop}%
	\bibitem [{\citenamefont {Loss}\ and\ \citenamefont
		{DiVincenzo}(1998)}]{Loss1998}%
	\BibitemOpen
	\bibfield  {author} {\bibinfo {author} {\bibfnamefont {D.}~\bibnamefont
			{Loss}}\ and\ \bibinfo {author} {\bibfnamefont {D.~P.}~\bibnamefont
			{DiVincenzo}},\ }\href@noop {} {\bibfield  {journal} {\bibinfo  {journal}
			{Phys. Rev. A}\ }\textbf {\bibinfo {volume} {57}},\ \bibinfo {pages} {120}
		(\bibinfo {year} {1998})}\BibitemShut {NoStop}%
	\bibitem [{\citenamefont {Burkard}\ \emph {et~al.}(1999)\citenamefont
		{Burkard}, \citenamefont {Loss},\ and\ \citenamefont
		{DiVincenzo}}]{Burkard1999}%
	\BibitemOpen
	\bibfield  {author} {\bibinfo {author} {\bibfnamefont {G.}~\bibnamefont
			{Burkard}}, \bibinfo {author} {\bibfnamefont {D.}~\bibnamefont {Loss}}, \
		and\ \bibinfo {author} {\bibfnamefont {D.~P.}~\bibnamefont {DiVincenzo}},\
	}\href@noop {} {\bibfield  {journal} {\bibinfo  {journal} {Phys. Rev. B}\
		}\textbf {\bibinfo {volume} {59}},\ \bibinfo {pages} {2070} (\bibinfo {year}
		{1999})}\BibitemShut {NoStop}%
	\bibitem [{\citenamefont {Bose}(2003)}]{Bose2003}%
	\BibitemOpen
	\bibfield  {author} {\bibinfo {author} {\bibfnamefont {S.}~\bibnamefont
			{Bose}},\ }\href@noop {} {\bibfield  {journal} {\bibinfo  {journal} {Phys.
				Rev. Lett.}\ }\textbf {\bibinfo {volume} {91}},\ \bibinfo {pages} {207901}
		(\bibinfo {year} {2003})}\BibitemShut {NoStop}%
	\bibitem [{\citenamefont {Das}\ and\ \citenamefont
		{Chakrabarti}(2008)}]{Das2008}%
	\BibitemOpen
	\bibfield  {author} {\bibinfo {author} {\bibfnamefont {A.}~\bibnamefont
			{Das}}\ and\ \bibinfo {author} {\bibfnamefont {B.~K.}\ \bibnamefont
			{Chakrabarti}},\ }\href@noop {} {\bibfield  {journal} {\bibinfo  {journal}
			{Rev. Mod. Phys.}\ }\textbf {\bibinfo {volume} {80}},\ \bibinfo {pages}
		{1061} (\bibinfo {year} {2008})}\BibitemShut {NoStop}%
	\bibitem [{\citenamefont {Nakazato}\ \emph {et~al.}(2003)\citenamefont
		{Nakazato}, \citenamefont {Takazawa},\ and\ \citenamefont
		{Yuasa}}]{Hiromichi}%
	\BibitemOpen
	\bibfield  {author} {\bibinfo {author} {\bibfnamefont {H.}~\bibnamefont
			{Nakazato}}, \bibinfo {author} {\bibfnamefont {T.}~\bibnamefont {Takazawa}},
		\ and\ \bibinfo {author} {\bibfnamefont {K.}~\bibnamefont {Yuasa}},\ }\href
	{\doibase 10.1103/PhysRevLett.90.060401} {\bibfield  {journal} {\bibinfo
			{journal} {Phys. Rev. Lett.}\ }\textbf {\bibinfo {volume} {90}},\ \bibinfo
		{pages} {060401} (\bibinfo {year} {2003})}\BibitemShut {NoStop}%
	\bibitem [{\citenamefont {Wu}\ \emph {et~al.}(2004)\citenamefont {Wu},
		\citenamefont {Lidar},\ and\ \citenamefont
		{Schneider}}]{Wu_entanglement_generation}%
	\BibitemOpen
	\bibfield  {author} {\bibinfo {author} {\bibfnamefont {L.-A.}\ \bibnamefont
			{Wu}}, \bibinfo {author} {\bibfnamefont {D.~A.}\ \bibnamefont {Lidar}}, \
		and\ \bibinfo {author} {\bibfnamefont {S.}~\bibnamefont {Schneider}},\ }\href
	{\doibase 10.1103/PhysRevA.70.032322} {\bibfield  {journal} {\bibinfo
			{journal} {Phys. Rev. A}\ }\textbf {\bibinfo {volume} {70}},\ \bibinfo
		{pages} {032322} (\bibinfo {year} {2004})}\BibitemShut {NoStop}%
	\bibitem [{\citenamefont {Glazman}\ and\ \citenamefont
		{Ioselevich}(1988)}]{Glazman_Polaron}%
	\BibitemOpen
	\bibfield  {author} {\bibinfo {author} {\bibfnamefont {L.~I.}\ \bibnamefont
			{Glazman}}\ and\ \bibinfo {author} {\bibfnamefont {A.~S.}\ \bibnamefont
			{Ioselevich}},\ }\href@noop {} {\bibfield  {journal} {\bibinfo  {journal}
			{JETP Lett.}\ }\textbf {\bibinfo {volume} {47}},\ \bibinfo {pages} {547}
		(\bibinfo {year} {1988})}\BibitemShut {NoStop}%
	\bibitem [{\citenamefont {Grusdt}\ \emph {et~al.}(2018)\citenamefont {Grusdt},
		\citenamefont {K\'anasz-Nagy}, \citenamefont {Bohrdt}, \citenamefont {Chiu},
		\citenamefont {Ji}, \citenamefont {Greiner}, \citenamefont {Greif},\ and\
		\citenamefont {Demler}}]{parton-polaron}%
	\BibitemOpen
	\bibfield  {author} {\bibinfo {author} {\bibfnamefont {F.}~\bibnamefont
			{Grusdt}}, \bibinfo {author} {\bibfnamefont {M.}~\bibnamefont
			{K\'anasz-Nagy}}, \bibinfo {author} {\bibfnamefont {A.}~\bibnamefont
			{Bohrdt}}, \bibinfo {author} {\bibfnamefont {C.~S.}\ \bibnamefont {Chiu}},
		\bibinfo {author} {\bibfnamefont {G.}~\bibnamefont {Ji}}, \bibinfo {author}
		{\bibfnamefont {M.}~\bibnamefont {Greiner}}, \bibinfo {author} {\bibfnamefont
			{D.}~\bibnamefont {Greif}}, \ and\ \bibinfo {author} {\bibfnamefont
			{E.}~\bibnamefont {Demler}},\ }\href {\doibase 10.1103/PhysRevX.8.011046}
	{\bibfield  {journal} {\bibinfo  {journal} {Phys. Rev. X}\ }\textbf {\bibinfo
			{volume} {8}},\ \bibinfo {pages} {011046} (\bibinfo {year}
		{2018})}\BibitemShut {NoStop}%
	\bibitem [{\citenamefont {Prilepsky}\ \emph {et~al.}(2006)\citenamefont
		{Prilepsky}, \citenamefont {Kovalev}, \citenamefont {Johansson},\ and\
		\citenamefont {Kivshar}}]{polaron-classical-1}%
	\BibitemOpen
	\bibfield  {author} {\bibinfo {author} {\bibfnamefont {J.~E.}\ \bibnamefont
			{Prilepsky}}, \bibinfo {author} {\bibfnamefont {A.~S.}\ \bibnamefont
			{Kovalev}}, \bibinfo {author} {\bibfnamefont {M.}~\bibnamefont {Johansson}},
		\ and\ \bibinfo {author} {\bibfnamefont {Y.~S.}\ \bibnamefont {Kivshar}},\
	}\href {\doibase 10.1103/PhysRevB.74.132404} {\bibfield  {journal} {\bibinfo
			{journal} {Phys. Rev. B}\ }\textbf {\bibinfo {volume} {74}},\ \bibinfo
		{pages} {132404} (\bibinfo {year} {2006})}\BibitemShut {NoStop}%
	\bibitem [{\citenamefont {Nakano}\ and\ \citenamefont
		{Takahashi}(2003)}]{polaron-quantum-1}%
	\BibitemOpen
	\bibfield  {author} {\bibinfo {author} {\bibfnamefont {H.}~\bibnamefont
			{Nakano}}\ and\ \bibinfo {author} {\bibfnamefont {Y.}~\bibnamefont
			{Takahashi}},\ }\href {\doibase 10.1143/JPSJ.72.1191} {\bibfield  {journal}
		{\bibinfo  {journal} {Journal of the Physical Society of Japan}\ }\textbf
		{\bibinfo {volume} {72}},\ \bibinfo {pages} {1191} (\bibinfo {year}
		{2003})}\BibitemShut {NoStop}%
	\bibitem [{\citenamefont {Hayn}\ \emph {et~al.}(1996)\citenamefont {Hayn},
		\citenamefont {Barabanov}, \citenamefont {Schulenburg},\ and\ \citenamefont
		{Richter}}]{Barabanov-Spin-Polaron}%
	\BibitemOpen
	\bibfield  {author} {\bibinfo {author} {\bibfnamefont {R.}~\bibnamefont
			{Hayn}}, \bibinfo {author} {\bibfnamefont {A.~F.}\ \bibnamefont {Barabanov}},
		\bibinfo {author} {\bibfnamefont {J.}~\bibnamefont {Schulenburg}}, \ and\
		\bibinfo {author} {\bibfnamefont {J.}~\bibnamefont {Richter}},\ }\href@noop
	{} {\bibfield  {journal} {\bibinfo  {journal} {Phys. Rev. B}\ }\textbf
		{\bibinfo {volume} {53}},\ \bibinfo {pages} {11714} (\bibinfo {year}
		{1996})}\BibitemShut {NoStop}%
	\bibitem [{\citenamefont {Li}\ \emph {et~al.}(2011)\citenamefont {Li},
		\citenamefont {Wu}, \citenamefont {Wang},\ and\ \citenamefont
		{Yang}}]{Li2011}%
	\BibitemOpen
	\bibfield  {author} {\bibinfo {author} {\bibfnamefont {Y.}~\bibnamefont
			{Li}}, \bibinfo {author} {\bibfnamefont {L.-A.}\ \bibnamefont {Wu}}, \bibinfo
		{author} {\bibfnamefont {Y.-D.}\ \bibnamefont {Wang}}, \ and\ \bibinfo
		{author} {\bibfnamefont {L.-P.}\ \bibnamefont {Yang}},\ }\href {\doibase
		10.1103/PhysRevB.84.094502} {\bibfield  {journal} {\bibinfo  {journal} {Phys.
				Rev. B}\ }\textbf {\bibinfo {volume} {84}},\ \bibinfo {pages} {094502}
		(\bibinfo {year} {2011})}\BibitemShut {NoStop}%
	\bibitem [{\citenamefont {Lee}\ and\ \citenamefont
		{Yang}(1952)}]{imag-magn-field-1952}%
	\BibitemOpen
	\bibfield  {author} {\bibinfo {author} {\bibfnamefont {T.~D.}\ \bibnamefont
			{Lee}}\ and\ \bibinfo {author} {\bibfnamefont {C.~N.}\ \bibnamefont {Yang}},\
	}\href {\doibase 10.1103/PhysRev.87.410} {\bibfield  {journal} {\bibinfo
			{journal} {Phys. Rev.}\ }\textbf {\bibinfo {volume} {87}},\ \bibinfo {pages}
		{410} (\bibinfo {year} {1952})}\BibitemShut {NoStop}%
	\bibitem [{\citenamefont {McCoy}\ and\ \citenamefont
		{Wu}(1967)}]{imag-magn-field-1967}%
	\BibitemOpen
	\bibfield  {author} {\bibinfo {author} {\bibfnamefont {B.~M.}\ \bibnamefont
			{McCoy}}\ and\ \bibinfo {author} {\bibfnamefont {T.~T.}\ \bibnamefont {Wu}},\
	}\href {\doibase 10.1103/PhysRev.155.438} {\bibfield  {journal} {\bibinfo
			{journal} {Phys. Rev.}\ }\textbf {\bibinfo {volume} {155}},\ \bibinfo {pages}
		{438} (\bibinfo {year} {1967})}\BibitemShut {NoStop}%
	\bibitem [{\citenamefont {Azcoiti}\ \emph {et~al.}(2017)\citenamefont
		{Azcoiti}, \citenamefont {Di~Carlo}, \citenamefont {Follana},\ and\
		\citenamefont {Royo-Amondarain}}]{imag-magn-field-2017}%
	\BibitemOpen
	\bibfield  {author} {\bibinfo {author} {\bibfnamefont {V.}~\bibnamefont
			{Azcoiti}}, \bibinfo {author} {\bibfnamefont {G.}~\bibnamefont {Di~Carlo}},
		\bibinfo {author} {\bibfnamefont {E.}~\bibnamefont {Follana}}, \ and\
		\bibinfo {author} {\bibfnamefont {E.}~\bibnamefont {Royo-Amondarain}},\
	}\href {\doibase 10.1103/PhysRevE.96.032114} {\bibfield  {journal} {\bibinfo
			{journal} {Phys. Rev. E}\ }\textbf {\bibinfo {volume} {96}},\ \bibinfo
		{pages} {032114} (\bibinfo {year} {2017})}\BibitemShut {NoStop}%
	\bibitem [{\citenamefont {Lee}\ and\ \citenamefont
		{Chan}(2014)}]{non-herm-magnetism-PhysRevX.4.041001}%
	\BibitemOpen
	\bibfield  {author} {\bibinfo {author} {\bibfnamefont {T.~E.}\ \bibnamefont
			{Lee}}\ and\ \bibinfo {author} {\bibfnamefont {C.-K.}\ \bibnamefont {Chan}},\
	}\href {\doibase 10.1103/PhysRevX.4.041001} {\bibfield  {journal} {\bibinfo
			{journal} {Phys. Rev. X}\ }\textbf {\bibinfo {volume} {4}},\ \bibinfo {pages}
		{041001} (\bibinfo {year} {2014})}\BibitemShut {NoStop}%
	\bibitem [{\citenamefont {Dattoli}\ \emph {et~al.}(1990)\citenamefont
		{Dattoli}, \citenamefont {Torre},\ and\ \citenamefont
		{Mignani}}]{non-herm-evol-two-level-quant-sys-PhysRevA.42.1467}%
	\BibitemOpen
	\bibfield  {author} {\bibinfo {author} {\bibfnamefont {G.}~\bibnamefont
			{Dattoli}}, \bibinfo {author} {\bibfnamefont {A.}~\bibnamefont {Torre}}, \
		and\ \bibinfo {author} {\bibfnamefont {R.}~\bibnamefont {Mignani}},\ }\href
	{\doibase 10.1103/PhysRevA.42.1467} {\bibfield  {journal} {\bibinfo
			{journal} {Phys. Rev. A}\ }\textbf {\bibinfo {volume} {42}},\ \bibinfo
		{pages} {1467} (\bibinfo {year} {1990})}\BibitemShut {NoStop}%
	\bibitem [{\citenamefont {Edelman}\ \emph {et~al.}(1994)\citenamefont
		{Edelman}, \citenamefont {Kostlan},\ and\ \citenamefont
		{Shub}}]{real_eigenvalues}%
	\BibitemOpen
	\bibfield  {author} {\bibinfo {author} {\bibfnamefont {A.}~\bibnamefont
			{Edelman}}, \bibinfo {author} {\bibfnamefont {E.}~\bibnamefont {Kostlan}}, \
		and\ \bibinfo {author} {\bibfnamefont {M.}~\bibnamefont {Shub}},\ }\href@noop
	{} {\bibfield  {journal} {\bibinfo  {journal} {Journal of the American
				Mathematical Society}\ }\textbf {\bibinfo {volume} {7}},\ \bibinfo {pages}
		{247} (\bibinfo {year} {1994})}\BibitemShut {NoStop}%
	\bibitem [{\citenamefont {Wu}\ \emph {et~al.}(2005)\citenamefont {Wu},
		\citenamefont {Zanardi},\ and\ \citenamefont {Lidar}}]{Wu2005}%
	\BibitemOpen
	\bibfield  {author} {\bibinfo {author} {\bibfnamefont {L.-A.}\ \bibnamefont
			{Wu}}, \bibinfo {author} {\bibfnamefont {P.}~\bibnamefont {Zanardi}}, \ and\
		\bibinfo {author} {\bibfnamefont {D.~A.}\ \bibnamefont {Lidar}},\ }\href@noop
	{} {\bibfield  {journal} {\bibinfo  {journal} {Phys. Rev. Lett.}\ }\textbf
		{\bibinfo {volume} {95}},\ \bibinfo {pages} {130501} (\bibinfo {year}
		{2005})}\BibitemShut {NoStop}%
	\bibitem [{\citenamefont {Gily{\'{e}}n}\ \emph {et~al.}(2016)\citenamefont
		{Gily{\'{e}}n}, \citenamefont {Kiss},\ and\ \citenamefont
		{Jex}}]{exponential-sens}%
	\BibitemOpen
	\bibfield  {author} {\bibinfo {author} {\bibfnamefont {A.}~\bibnamefont
			{Gily{\'{e}}n}}, \bibinfo {author} {\bibfnamefont {T.}~\bibnamefont {Kiss}},
		\ and\ \bibinfo {author} {\bibfnamefont {I.}~\bibnamefont {Jex}},\
	}\href@noop {} {\bibfield  {journal} {\bibinfo  {journal} {Scientific
				Reports}\ }\textbf {\bibinfo {volume} {6}} (\bibinfo {year}
		{2016})}\BibitemShut {NoStop}%
	\bibitem [{\citenamefont {Zhu}\ \emph {et~al.}(2019)\citenamefont {Zhu},
		\citenamefont {K\'alm\'an}, \citenamefont {Wang}, \citenamefont {Xiao},
		\citenamefont {Qu}, \citenamefont {Zhan}, \citenamefont {Bian}, \citenamefont
		{Kiss},\ and\ \citenamefont {Xue}}]{Measur_indused_orthogonalization}%
	\BibitemOpen
	\bibfield  {author} {\bibinfo {author} {\bibfnamefont {G.}~\bibnamefont
			{Zhu}}, \bibinfo {author} {\bibfnamefont {O.}~\bibnamefont {K\'alm\'an}},
		\bibinfo {author} {\bibfnamefont {K.}~\bibnamefont {Wang}}, \bibinfo {author}
		{\bibfnamefont {L.}~\bibnamefont {Xiao}}, \bibinfo {author} {\bibfnamefont
			{D.}~\bibnamefont {Qu}}, \bibinfo {author} {\bibfnamefont {X.}~\bibnamefont
			{Zhan}}, \bibinfo {author} {\bibfnamefont {Z.}~\bibnamefont {Bian}}, \bibinfo
		{author} {\bibfnamefont {T.}~\bibnamefont {Kiss}}, \ and\ \bibinfo {author}
		{\bibfnamefont {P.}~\bibnamefont {Xue}},\ }\href@noop {} {\bibfield
		{journal} {\bibinfo  {journal} {Phys. Rev. A}\ }\textbf {\bibinfo {volume}
			{100}},\ \bibinfo {pages} {052307} (\bibinfo {year} {2019})}\BibitemShut
	{NoStop}%
	\bibitem [{\citenamefont {Pyshkin}\ \emph {et~al.}(2018)\citenamefont
		{Pyshkin}, \citenamefont {Sherman}, \citenamefont {You},\ and\ \citenamefont
		{Wu}}]{Pyshkin_Spin_cutting_NJP}%
	\BibitemOpen
	\bibfield  {author} {\bibinfo {author} {\bibfnamefont {P.~V.}\ \bibnamefont
			{Pyshkin}}, \bibinfo {author} {\bibfnamefont {E.~Y.}\ \bibnamefont
			{Sherman}}, \bibinfo {author} {\bibfnamefont {J.~Q.}\ \bibnamefont {You}}, \
		and\ \bibinfo {author} {\bibfnamefont {L.-A.}\ \bibnamefont {Wu}},\
	}\href@noop {} {\bibfield  {journal} {\bibinfo  {journal} {New Journal of
				Physics}\ }\textbf {\bibinfo {volume} {20}},\ \bibinfo {pages} {105006}
		(\bibinfo {year} {2018})}\BibitemShut {NoStop}%
	\bibitem [{\citenamefont {Pyshkin}\ \emph {et~al.}(2019)\citenamefont
		{Pyshkin}, \citenamefont {Sherman},\ and\ \citenamefont
		{Wu}}]{Spin_Chain_local_cut}%
	\BibitemOpen
	\bibfield  {author} {\bibinfo {author} {\bibfnamefont {P.~V.}\ \bibnamefont
			{Pyshkin}}, \bibinfo {author} {\bibfnamefont {E.~Y.}\ \bibnamefont
			{Sherman}}, \ and\ \bibinfo {author} {\bibfnamefont {L.-A.}\ \bibnamefont
			{Wu}},\ }\href@noop {} {\bibfield  {journal} {\bibinfo  {journal} {Phys. Rev.
				A}\ }\textbf {\bibinfo {volume} {100}},\ \bibinfo {pages} {063401} (\bibinfo
		{year} {2019})}\BibitemShut {NoStop}%
	\bibitem [{\citenamefont {Bae}\ and\ \citenamefont
		{Kwek}(2015)}]{discrimination_Bae_2015}%
	\BibitemOpen
	\bibfield  {author} {\bibinfo {author} {\bibfnamefont {J.}~\bibnamefont
			{Bae}}\ and\ \bibinfo {author} {\bibfnamefont {L.-C.}\ \bibnamefont {Kwek}},\
	}\href@noop {} {\bibfield  {journal} {\bibinfo  {journal} {Journal of Physics
				A: Mathematical and Theoretical}\ }\textbf {\bibinfo {volume} {48}},\
		\bibinfo {pages} {083001} (\bibinfo {year} {2015})}\BibitemShut {NoStop}%
\end{thebibliography}

%merlin.mbs apsrev4-1.bst 2010-07-25 4.21a (PWD, AO, DPC) hacked
%Control: key (0)
%Control: author (72) initials jnrlst
%Control: editor formatted (1) identically to author
%Control: production of article title (-1) disabled
%Control: page (0) single
%Control: year (1) truncated
%Control: production of eprint (0) enabled
%

\end{document}